\documentclass{jetpl}
\twocolumn

\usepackage{amsmath,amssymb,color,pstricks,bm,alltt}

\usepackage[colorlinks,plainpages=false,linkcolor=black,urlcolor=blue,
citecolor=black,pdfpagemode=UseNone,pdfstartview=FitBH]{hyperref}

\begin{document}

%%% article in English
\lat

%%% article title
\title{On the origin of the shallow and ``replica'' bands in FeSe
monolayer superconductors}

%%% article title - for colontitle (at the top of the page)
\rtitle{On the origin of the shallow and ``replica'' bands in FeSe
monolayer superconductors}

%%% article title - for table of contents (usually identical with \title)
\sodtitle{On the origin of the shallow and ``replica'' bands in FeSe
monolayer superconductors}

%%% author(s) ( + e-mail)
\author{$^a$I.\ A.\ Nekrasov\thanks{E-mail: nekrasov@iep.uran.ru},
$^a$N.\ S.\ Pavlov\thanks{E-mail: pavlov@iep.uran.ru},
$^{a,b}$M.\ V.\ Sadovskii\thanks{E-mail: sadovski@iep.uran.ru}
}

%%% author(s) - for colontitle (at the top of the page)
\rauthor{I.\ A.\ Nekrasov, N.\ S.\ Pavlov, M.\ V.\ Sadovskii}

%%% author(s) - for table of contents
\sodauthor{I.\ A.\ Nekrasov, N.\ S.\ Pavlov, M.\ V.\ Sadovskii }

%%% author(s) - for table of contents
\sodauthor{I.\ A.\ Nekrasov, N.\ S.\ Pavlov, M.\ V.\ Sadovskii}

%%% author's address(es)
\address{$^a$Institute of Electrophysics, Russian Academy of Sciences,
Ural Branch, Amundsen str. 106,  Ekaterinburg 620016, Russia\\
$^b$ M.N. Mikheev Institute of Metal Physics, Russian Academy of Sciences, Ural Branch,
S. Kovalevskaya str. 18, Ekaterinburg 620290, Russia
}

%%% dates of submition & resubmition (if submitted once, second argument is *)
%\dates{}{*}

\abstract{We compare electronic structures of single FeSe layer films on
SrTiO$_3$ substrate (FeSe/STO) and K$_x$Fe$_{2-y}$Se$_{2}$ superconductors
obtained from extensive LDA and LDA+DMFT calculations with the results of
ARPES experiments.
It is demonstrated that correlation effects on Fe-3d states are sufficient
in principle to explain the formation of the shallow electron -- like bands at
the M(X)-point. However, in FeSe/STO these effects alone are apparently
insufficient for the simultaneous elimination of the hole -- like Fermi surface
around the $\Gamma$-point which is not observed in ARPES experiments.
Detailed comparison of ARPES detected and calculated quasiparticle bands shows
reasonable agreement between theory and experiment. Analysis of the bands
with respect to their origin and orbital composition shows, that
for FeSe/STO system the experimentally observed ``replica'' quasiparticle band
at the M-point (usually attributed to forward scattering interactions with
optical phonons in SrTiO$_3$ substrate) can be reasonably understood just as the
LDA calculated Fe-3d$_{xy}$ band, renormalized by electronic correlations.
The only manifestation of the substrate reduces to lifting the degeneracy
between Fe-3d$_{xz}$ and Fe-3d$_{yz}$ bands in the vicinity of M-point.
For the case of K$_x$Fe$_{2-y}$Se$_{2}$ most bands
observed in ARPES can also be understood as correlation renormalized Fe-3d LDA
calculated bands, with overall semi -- quantitative agreement with LDA+DMFT
calculations.}


\maketitle

\section{Introduction}
The discovery of high-temperature superconductivity in iron pnictides
(see reviews~\cite{Sad_08, Hoso_09, John, MazKor, Stew, Kord_12} was almost
immediately followed by the observation of rather low temperature ($T_c\sim$ 8K)
superconductivity in iron {\em chalcogenide} FeSe, with
electronic structure quite similar to that of iron pnictides
(see review~\cite{FeSe}).

Further success in creation of {\em intercalated} FeSe based systems with rather
high $T_c\sim$ 30-40K (see review in~\cite{Maziopa,Sad_16}) quickly made them
popular objects of investigations because of their different electronic
structure~\cite{JMMM, JTLRev}.

Most impressive results were achieved with the growth of epitaxial films of
single FeSe monolayer on 001 plane of Sr(Ba)TiO$_3$ (STO) substrate
with record breaking $T_c$ values in the range of 65--85 K~\cite{FeSe_STO1,FeSe_BTO} (or probably up to 100 K~\cite{FeSe_STO2}).
The general theoretical and experimental situation in these rapidly developing
field of research was described in recent reviews~\cite{Sad_16,FeSe1UC_rev}.

ARPES measurements~\cite{FeSe_STO_arpes14} in FeSe/STO monolayer system
demonstrated rather unusual band structure, characterized by the absence of
hole -- like bands at the center of Brillouin zone ($\Gamma$-point), with rather
shallow electronic band at the M-point with very low Fermi energies of the order
of 50 meV, accompanied by the formation of ``replica'' of this band about 100 meV
below in energy.
Similar unusually shallow bands were also observed at
X-point in ARPES experiments on intercalated  K$_x$Fe$_{2-y}$Se$_2$ system~\cite{KFeSe_arpes16}.

The existence of such peculiar bands rises many serious theoretical questions~\cite{Sad_16}, such as probable considerable role of non adiabatic interactions~\cite{Gork_1,Gork_2} and the possibility of observation of BCS-BEC crossover effects
in these systems. In particular, the formation of the ``replica'' band in
FeSe/STO is widely interpreted as being due to inteaction with high -- energy
($\sim$ 100 meV) optical phonons of Sr(Ba)TiO$_3$ substrate~\cite{FeSe_STO_arpes14} with some important conclusions on the possible role of
these interactions for the significant enhancement of $T_c$ in this system~\cite{Gork_1,Gork_2,Rade_1,Rade_2}.

Further in this paper we compare the ARPES detected quasiparticle bands for
FeSe/STO and K$_x$Fe$_{2-y}$Se$_{2}$ and compare them with the results of
our LDA+DMFT calculations for these systems as well as for the  isolated FeSe
layer,  together with the  analysis of initial LDA bands~\cite{Nekrasov_FeSe}.
Interaction parameters of the Hubbard model in LDA+DMFT were taken $U$=5.0 eV,
$J$=0.9 eV for FeSe and FeSe/STO and $U$=3.75 eV, $J$=0.56 eV for
KFe$_2$Se$_2$ (see the Supplemental Material~\cite{Suppl}, where we present
further computational details).

\section{FeSe/STO system}
In Fig.~1 we compare the theoretical LDA+DMFT results on panels (a,d,e,h)
with experimental ARPES data~\cite{FeSe_STO_arpes14} on panels (b,c,f,g).
LDA+DMFT spectral function maps of isolated FeSe monolayer are shown in Fig.~1(a) and
Fig.~1(d) at $\Gamma$ and M points respectively. For FeSe/STO LDA+DMFT spectral
function maps are shown on (e), (h) panels at $\Gamma$ and M points.
The obtained LDA bandwidth of Fe-3d band in isolated FeSe monolayer is 5.2 eV, which is
much larger than 4.3 eV that obtained for FeSe/STO. This is due to the
lattice constant $a$ expanded from $a=3.765$~\AA\ to $a=3.901$~\AA~in going from isolated
FeSe monolayer to FeSe/STO. Thus for the same interaction strength and doping
levels LDA+DMFT calculations show substantially different band narrowing due to
correlation effects: a factor of 1.5 in isolated FeSe monolayer (same as bulk FeSe) and
a factor of 3 in FeSe/STO.
Thus $ceteris~paribus$ FeSe/STO system is more correlated as compared with the
bulk FeSe or isolated FeSe layer.

Most of features observed in the ARPES data (Fig.~1, panels (f),(g)) can be
identified with our calculated LDA+DMFT spectral function maps
(Fig.~1, panels (e),(h)).
The experimental quasiparticle bands around M-point marked by $A$, $B$ and $C$
(Fig.~1(g,h)) correspond mainly to Fe-3d$_{xz}$ and Fe-3d$_{yz}$ states,
while the $A'$ and $B'$ quasiparticle bands have predominantly Fe-3d$_{xy}$
character. As we noted above the appearance of $A'$ band in FeSe/STO is
usually attributed to forward scattering interaction with 100 meV optical
phonon of STO substrate~\cite{FeSe_STO_arpes14,Gork_1,Gork_2,Rade_1,Rade_2}.
However, our calculations show that $A'$ band is most probably of purely
electronic nature. Some puzzling behavior of this band can be explained by
difficulties of experimental observations of the Fe-3d$_{xy}$ states near
M-point (see Refs.~\cite{KFeSe_arpes16,FeSe_arpes_dxy} and references therein,
as well as discussion in Ref.~\cite{NaFeSe} in the context of NaFeAs
compound).

Thus, in FeSe/STO we observe the overall agreement between LDA+DMFT results
(Fig.~1(h))  and ARPES data~\cite{FeSe_STO_arpes14} (Fig.~1(g)) on
semi-quantitative level with respect to relative positions of quasiparticle bands.
Let us also note that the Fermi surfaces formed by the $A$ and $A'$ bands in our LDA+DMFT calculations
are nearly the same as the Fermi surface observed at M-point by ARPES.

The shallow band at M-point originates from LDA Fe-3d$_{xz}$ and Fe-3d$_{yz}$
bands (see also Fig.~2, right panel) compressed by electronic correlations.
In the hope of achieving the better agreement with experiments we also examined
the reasonable increase of Coulomb interaction within LDA+DMFT and the different doping levels,
but these have not produced the significant improvement of our results.

The $C$ quasiparticle band near M-point appeared because of the lifting of
degeneracy of Fe-3d$_{xz}$ and Fe-3d$_{yz}$ bands (in contrast to isolated FeSe
layer, see panel Fig.~1(d)). The origin of this band splitting is related to the
$z_{Se}$ height difference below and above Fe ions plane due to the presence of
interface with SrTiO$_3$ (see Supplemental Material~\cite{Suppl} for ion positions used in our calculations).

Actually, all quasiparticle bands in the vicinity of M-point can be well
represented as LDA bands compressed by a factor of 3 due to electronic
correlations.  This fact is clearly supported by LDA band structure shown on
the right panel of Fig.~2, where different bands are marked by letters identical
to those used in Fig.~1.

Near the M-point we can also observe the O-2p$_y$ band (in the energy interval
below -0.2 eV (Fig.~1(h)) originating from TiO$_2$ layer adjacent to FeSe.
Due to doping level used here this O-2p$_y$ band goes below the Fermi level in
contrast to LDA picture shown in Fig.~2 (on the right) where O-2p$_y$ band
crosses the Fermi level and forms hole pocket. This observation rules out
possible nesting effects which might be expected from LDA
results~\cite{Nekrasov_FeSe}.

Now let us discuss the bands around the $\Gamma$-point, which are presented on
panels (a,b,e,f) of Fig.~1. Here the situation is much simpler than in the case
of M-point. One can see here only two bands observed in the experiment (Fig.~1(f)).
The $D$ quasiparticle band has predominantly Fe-3d$_{xy}$ character, while the
$D'$ quasiparticle band originates from Fe-3d$_{3z^2-r^2}$ states.
Again the relative locations of LDA+DMFT calculated $D$ and $D'$ bands are quite
similar to the ARPES data.
\begin{figure*}[!ht]
\center{\includegraphics[width=.95\linewidth]{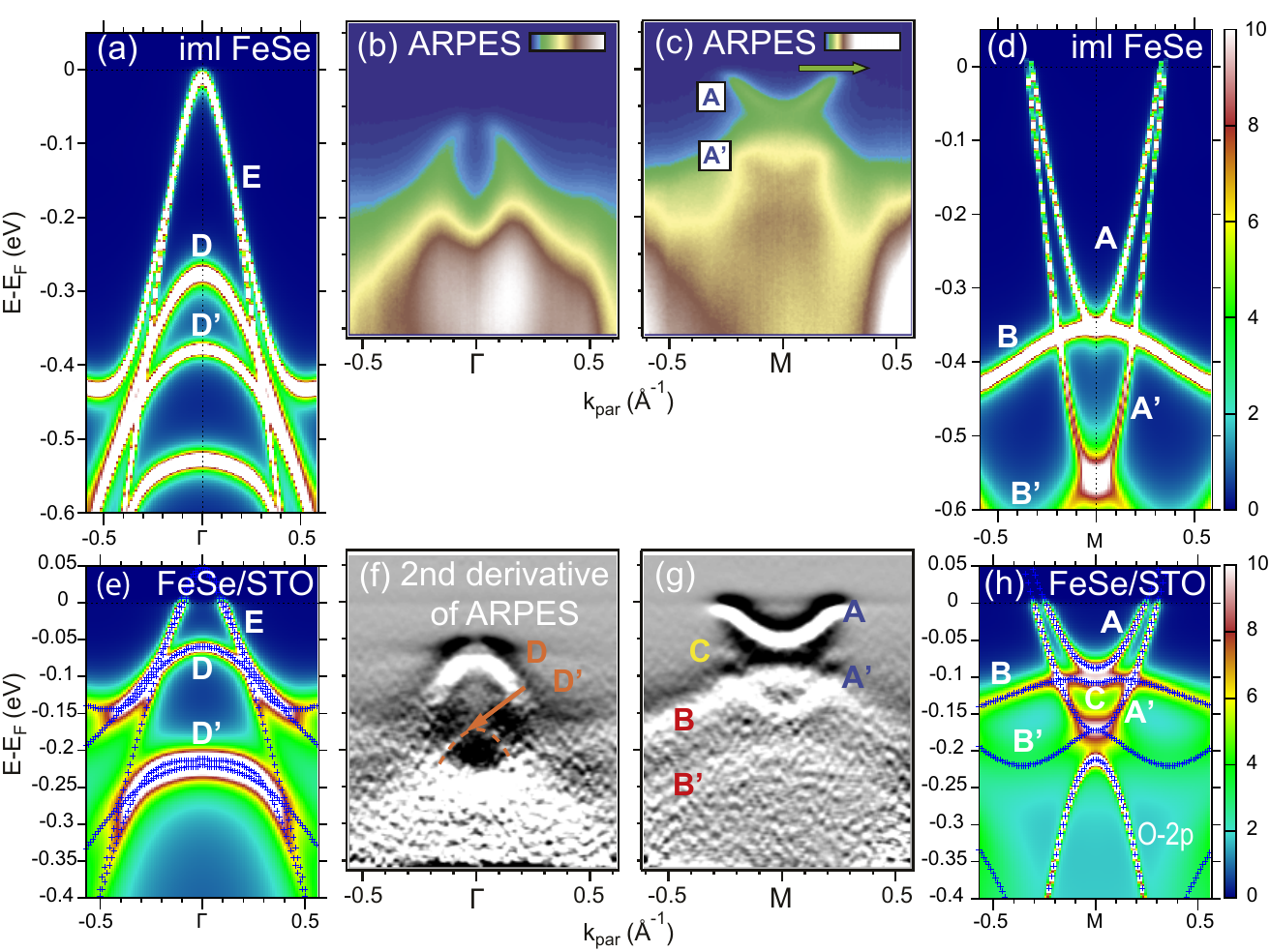}}
\caption{Fig. 1. (a), (d) panels -- LDA+DMFT spectral function maps of isolated FeSe
monolayer and (b),(c) -- experimental ARPES data around $\Gamma$ and M points
and (f), (g) corresponding second derivatives of ARPES data for
FeSe/STO~\cite{FeSe_STO_arpes14}; (e), (h) -- LDA+DMFT spectral function maps
with maxima shown with crosses for FeSe/STO. To mark similar features of
experimental and theoretical spectral function maps $A,B,C,D,E$ letters are
used (the same as in Fig.~2 for LDA bands). Fermi level is at zero energy.}
\label{fig1}
\end{figure*}
\begin{figure*}[!ht]
\center{\includegraphics[width=.7\textwidth]{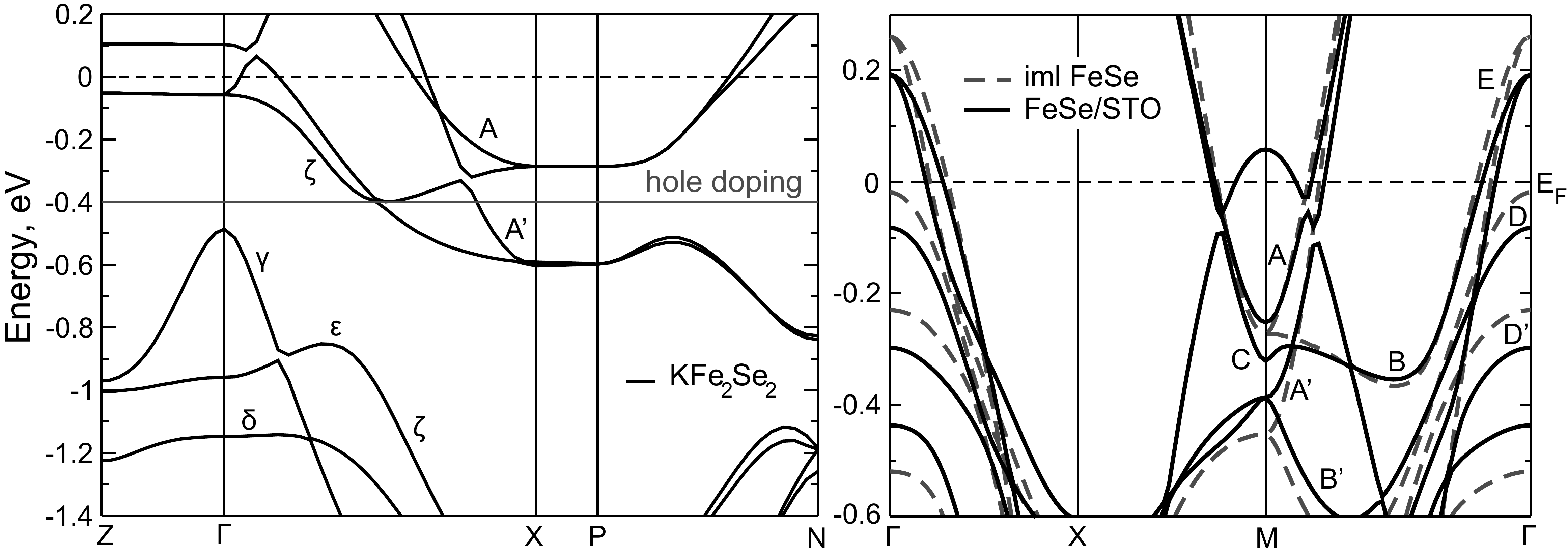}}
\caption{Fig. 2. LDA$'$ band dispersions of paramagnetic KFe$_2$Se$_2$ (left)
and LDA band dispersions of paramagnetic isolated FeSe monolayer (dashed line) and
paramagnetic FeSe/STO (solid line) (right). The letters designate bands in the
same way as in Fig.~1 and Fig.~3. The Fermi level $E_F$ is at zero energy.}
\label{fig2}
\end{figure*}

Main discrepancy of LDA+DMFT results and ARPES data is the $E$ band shown in
Fig.~1(e) which is not observed in the ARPES. This band corresponds to a
hybridized band of Fe-3d$_{xz}$, Fe-3d$_{yz}$ and Fe-3d$_{xy}$ states.
In principle some traces of this band can be guessed in the experimental
data of Fig.~1(f) around -0.17 eV and near the $k$-point 0.5. Surprisingly
these are missed in the discussion of Ref.~\cite{FeSe_STO_arpes14}.
Actually, the ARPES signal from $E$ band can be weakened because of sizable
Fe-3d$_{xy}$ contribution~\cite{KFeSe_arpes16,FeSe_arpes_dxy,NaFeSe} and
thus might be indistinguishable from $D$ band. Also one can imagine that for stronger band renormalization the $E$
band becomes more flat and might merge with $D$ band.
Detailed orbital resolved LDA+DMFT spectral function maps showing the intensity
of different orbital contributions are presented in Supplemental Material~\cite{Suppl}.
\begin{figure*}[!ht]
\center{\includegraphics[width=.85\linewidth]{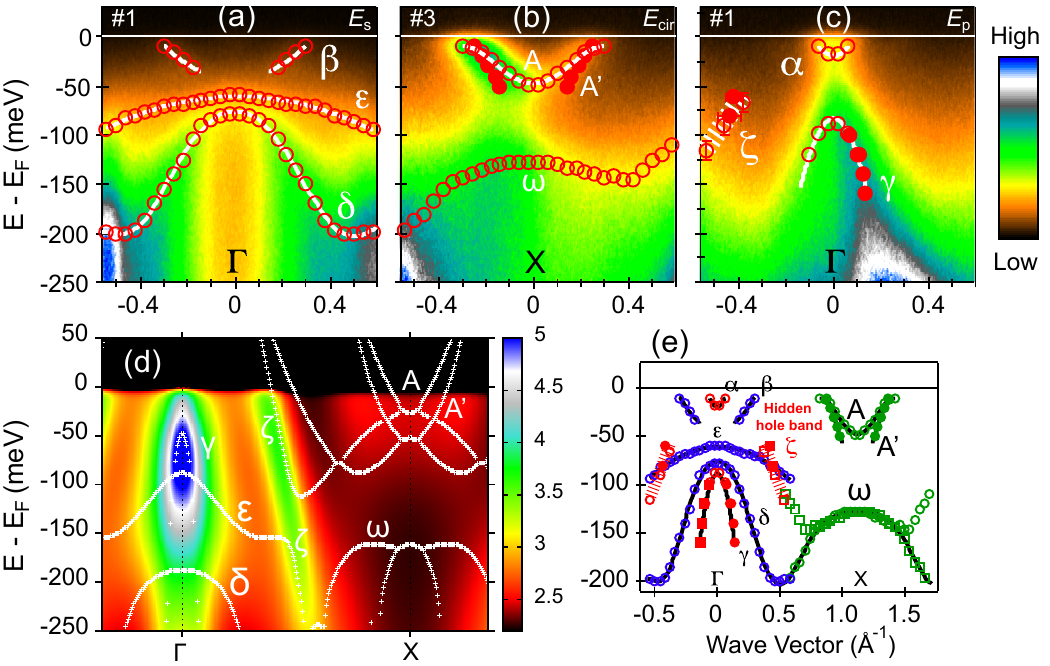}}
\caption{Fig. 3. Panels -- (a),(b),(c) ARPES data around $\Gamma$ and X-points
for K$_{0.62}$Fe$_{1.7}$Se$_2$~\cite{KFeSe_arpes16};
(d) -- LDA$'$+DMFT spectral function map with maxima shown by white crosses
for K$_{0.76}$Fe$_{1.72}$Se$_2$; (e) -- quasiparticle bands extracted from
ARPES~\cite{KFeSe_arpes16}. Bands of similar orbital character are marked with
Greek letters on all panels. Fermi level is at zero energy.}
\label{fig3}
\end{figure*}

\section{KFe$_2$Se$_2$ system}
In Fig.~3 we present the comparison of LDA+DMFT spectral function maps
(panel (d)) and ARPES data of Ref.~\cite{KFeSe_arpes16} (panels (a,b,c,e)) for K$_x$Fe$_{2-y}$Se$_{2}$.
Panels (a,b,c) of Fig.~3 correspond to different incident beam polarizations:
$E_s$ -- polarization in the plane parallel to the sample surface;
$E_p$ -- polarization in the plane normal to the sample surface;
$E_{cir}$ -- circular polarization. The use of different polarizations
allows one to distinguish contributions of bands with different symmetry
(see discussion in Ref.~\cite{KFeSe_arpes16,NaFeSe}). This fact is clearly seen in panels
(a,b,c) of Fig.~3 where different bands are marked with Greek letters.
In Fig.~3(e) we see the joint picture of all quasiparticle bands detected in
ARPES~\cite{KFeSe_arpes16} experiment.

Now we will try to explain the origin of the experimental bands and their
orbital composition on the basis of LDA$'$~\cite{LDA_prime_1,LDA_prime_2,LDA_prime_3,LDA_prime_4} calculations
for KFe$_2$Se$_2$ (Fig.~2, left panel) and LDA$'$+DMFT results (Fig.~3, panel (e)).
In our LDA$'$+DMFT calculations the $A$ quasiparticle band near X-point
corresponds to Fe-3d$_{xz}$ and Fe-3d$_{yz}$ states and the $A'$ quasiparticle
band near X-point is mainly formed by Fe-3d$_{xy}$ states.
These bands are denoted in the same way as on left panel of Fig.~2.
Thus the $A$ band corresponds to $\sim$ 50 meV shallow band typical for
FeSe monolayer materials. Its ``replica'' $A'$ band has Fe-3d$_{xy}$ symmetry
and is strongly suppressed in the experiments of Ref. \cite{KFeSe_arpes16}.
Actually, the authors of Ref.~\cite{KFeSe_arpes16} emphasized that they can not
obtain a signal from Fe-3d$_{xy}$ states. Thus both $A$ and $A'$
bands are just the renormalized LDA$'$ bands (compare with left panel of Fig.~2).
At about -0.15 eV at the X-point there is $\omega$ quasiparticle band which is
formed only due to self-energy effects.

Now we turn to bands around $\Gamma$-point. The $\varepsilon$ and $\delta$
bands are formed by Fe-3d$_{3z^2-r^2}$ states.
The $\varepsilon$ band is rather strongly modified in comparison
with the initial LDA$'$ $\varepsilon$ band (see Fig.~2, left panel),
while the $\delta$ band preserve the initial form rather well.
Energy location of $\varepsilon$ quasiparticle band agrees well for LDA$'$+DMFT
and ARPES results. However, the $\delta$ band is much lower in energy
in LDA$'$+DMFT. At the $\Gamma$-point the $\gamma$ band
(which is the hybridized band of Fe-3d$_{xz}$, Fe-3d$_{yz}$  and Fe-3d$_{xy}$
states) in LDA$'$+DMFT is above the $\varepsilon$ and $\delta$ bands in
contrast to ARPES data (Fig.~3(e)). This picture is somehow inherited from the
initial LDA$'$ band structure (Fig.~2, left). The $\zeta$ band
(Fig.~3(e)) consists in fact of two bands. The upper part (above 130 meV) of
this band is formed by Fe-3d$_{xz}$ and Fe-3d$_{yz}$ states, while the
lower part is formed by Fe-3d$_{3z^2-r^2}$ states. In ARPES experiments this
band is only partially observed around 80 meV (Fig.~3(e)), while
its lower part is not distinguished experimentally from $\omega$
band~\cite{KFeSe_arpes16}.

The overall agreement between ARPES and LDA$'$+DMFT results for
K$_{0.76}$Fe$_{1.72}$Se$_2$ system is rather satisfactory and allows one to
identify the orbital composition of different bands detected in the experiment.
However $\alpha$ and $\beta$ bands found in ARPES are not observed in our
LDA$'$+DMFT spectral function maps. More so there are no obvious candidates for
these bands within the LDA$'$ band structure (Fig.~2, left).
Thus the origin of experimentally observed $\alpha$ and $\beta$ quasiparticle
bands remains unclear.

\section{Conclusion}
Our results essentially allow the understanding of the origin of the shallow
band at the M-point in FeSe monolayer materials due to correlation effects
on Fe-3d states only.
The detailed analysis of ARPES detected quasiparticle bands and LDA+DMFT results
shows that this shallow band is formed just by the degenerate Fe-3d$_{xz}$ and
Fe-3d$_{yz}$ bands renormalized by correlations.
Moreover the so called ``replica'' band observed in ARPES for FeSe/STO can be
reasonably understood as the simple LDA renormalized Fe-3d$_{xy}$ band with no
reference to interactions with optical phonons of STO. The influence of STO
substrate in our calculations is reduced only to the removal of degeneracy of
Fe-3d$_{xz}$ and Fe-3d$_{yz}$ bands in the vicinity of M-point.
In the case of K$_x$Fe$_{2-y}$Se$_{2}$ most of ARPES detected bands can also be
expressed as correlation renormalized Fe-3d LDA bands.
Unfortunately correlation effects are unable to completely eliminate
the hole Fermi surface at the $\Gamma$-point, which is not observed in most
ARPES experiments on FeSe/STO system. Note, however, that
recently a small Fermi surface at the $\Gamma$-point was observed in ARPES
measurements on FeSe/STO at doping levels, corresponding to the highest values of
$T_c$ \cite{Shi_16}.

\section{Acknowledgements}
This work was done under the State contract (FASO) No. 0389-2014-0001 and
supported in part by RFBR grant No. 17-02-00015. NSP work was also supported by
the President of Russia grant for young scientists No. Mk-5957.2016.2.
The CT-QMC computations were performed at supercomputer ``Uran'' at the Institute of Mathematics and Mechanics UB RAS.

\newpage
\begin{center}
\Large
{\bf Supplemental Material to the article ``On the origin of the shallow and
``replica'' bands in FeSe monolayer superconductors''}
\end{center}
\setcounter{section}{0}
\setcounter{subsection}{0}

In this Supplement we provide computational details and crystallographic data
for FeSe  based systems under consideration. Also here we present orbital
resolved LDA+DMFT calculated quasipartical bands for these materials.

\section{Computation details}
The LDA$'$ calculations~\cite{CLDA,CLDA_long} of KFe$_2$Se$_2$ compound were
performed using the Linearized Muffin-Tin Orbitals method (LMTO)~\cite{LMTO}.
The electroning structures of FeSe monolayer and FeSe monolayer on SrTiO$_3$
substrate were calculated within FP-LAPW method~\cite{wien2k}.

For the DMFT part of LDA+DMFT calculations we employed CT-QMC impurity
solver~\cite{ctqmc,triqs}.
To define DMFT lattice problem for KFe$_2$Se$_2$ compound we used the full LDA
Hamiltonian (i.e. without any orbitals downfolding or projecting) same as in
Refs.~\cite{KFeSeLDADMFT1,KFeSeLDADMFT2}.
For isolated FeSe layer and FeSe/STO projection on Wannier functions was done
for Fe-3d and Se-4p states (isolated FeSe layer) and for Fe-3d, Se-4p states and
O-2p$_y$ states from TiO$_2$ layer adjacent to SrTiO$_3$ (FeSe/STO).
Standard  wien2wannier interface~\cite{wien2wannier} and wannier90 projecting
technique~\cite{wannier90} were applied to this end.

The DMFT(CT-QMC) computations were done at reciprocal temperature $\beta=40$
($\sim$290 K) with about 10$^8$ Monte-Carlo sweeps.
Interaction parameters of Hubbard model were taken $U$=5.0 eV, $J$=0.9 eV for
isolated FeSe and FeSe/STO and $U$=3.75 eV, $J$=0.56 eV for
KFe$_2$Se$_2$~\cite{KFeSe_arpes_UJ}. We employed the self-consistent
fully-localized limit definition of the double-counting
correction~\cite{CLDA_long}.
Thus computed values of Fe-3d occupancies and corresponding double-counting
energies are $E_{dc}=18.886$, $n_d=5.79$ (K$_{0.76}$Fe$_{1.72}$Se$_2$),
$E_{dc}=31.63$, $n_d=7.35$ (isolated FeSe layer), $E_{dc}=30.77$, $n_d=7.16$
(FeSe/STO).

The LDA+DMFT spectral function maps were obtained after analytic continuation
of the local self-energy $\Sigma(\omega)$ from Matsubara frequencies to the real
ones. To this end we have applied Pade approximant algorithm~\cite{pade} and
checked the results with the maximum entropy method~\cite{ME} for Green's
function G($\tau$).

\section{Crystal structure}
\subsection{FeSe, FeSe/STO}
The crystal structure of the bulk FeSe has tetragonal structure with the space
group $P$4/$nmm$ and lattice parameters $a=3.765$~\AA, $c=5.518$~\AA.
The experimentally obtained crystallographic positions are the following: Fe(2a)
(0.0, 0.0, 0.0), Se(2c) (0.0, 0.5, z$_{Se}$), z$_{Se}$=0.2343~\cite{FeSe_param}.
For LDA calculation of isolated FeSe layer the slab technique was applied with
these crystallographic parameters.

The FeSe/STO crystal structure was taken from LDA calculation with crystal
structure relaxation~\cite{FeSe_STO_param}. FeSe monolayer was placed on three
TiO$_2$-SrO layers to model the bulk SrTiO$_3$ substrate. The FeSe/STO slab
crystal structure parameters are $a=3.901$~\AA, Ti-Se distance $3.13$~\AA,
Fe-O distance $4.43$~\AA,  distance between  top (bottom) Se ion and the Fe
ions plane is 1.41~\AA\ (1.3~\AA). Atomic positions are:
Sr -- (0.5,0.5,-1.95~\AA), O -- (0.5,0,0), (0,0,-1.95~\AA), Ti -- (0,0,0).

LDA+DMFT calculations of FeSe/STO were performed for doping level of 0.2 electrons per Fe ion.

\subsection{KFe$_2$Se$_2$}
The ideal KFe$_2$Se$_2$ compound has tetragonal structure with the space group
$I$4/$mmm$ and lattice parameters $a=3.9136$~\AA~and $c=14.0367$~\AA.
The crystallographic positions are the following: K(2a)
(0.0, 0.0, 0.0), Fe(4d) (0.0, 0.5, 0.25), Se(4e) (0.0, 0.5, z$_{Se}$) with
z$_{Se}$=0.3539~\cite{KFeSe_cryst}.

Chemical composition K$_{0.76}$Fe$_{1.72}$Se$_2$ corresponds to the total number
of electrons 26.52 per unit cell (the stoichiometric compound has total number
of electrons equal to 29.0). Total number of electrons 26.52 per unit cell
corresponds to the doping level of 1.24 holes per Fe ion. This doping level was
taken for LDA$'$+DMFT calculations. Position of corresponding Fermi level at
about -0.4 eV is shown on left panel of Fig.~2 (main part of article).

\section{LDA+DMFT orbital resolved qusiparticle bands}
To show different Fe-3d orbitals contribution to LDA+DMFT spectral functions
of FeSe based systems under consideration we present here the corresponding
orbital resolved spectral function maps (Fig.~S~I,~II.).
In Fig.~S~I it is clearly seen that the qusiparticle bands of isolated FeSe
monolayer are well defined and have similar shape to the LDA bands except
correlation narrowing by the same constant factor for all bands.
The qusiparticle bands of FeSe/STO are more broad but still well defined.
The main contribution to spectral function near the Fermi level belongs to
Fe-3d$_{xz}$, Fe-3d$_{yz}$ and Fe-3d$_{xy}$ states both for the isolated FeSe
layer and FeSe/STO. The spectral function of
K$_{0.76}$Fe$_{1.72}$Se$_2$ is shown in Fig.~S~II.
Here the bands are strongly renormalized by
correlations not only by the constant scaling factor, but also because of band
shapes modifications in comparison to LDA bands. Since electronic correlations
are quite strong for K$_{0.76}$Fe$_{1.72}$Se$_2$ and bands are rather broadened
by lifetime effects we explicitly show here the spectral function maxima
positions by crosses.
Despite the difference of correlation effects in both systems one can conclude
that qusiparticle bands structures around the Fermi level are rather similar.
\begin{figure*}[!ht]
\center{\includegraphics[width=.58\linewidth]{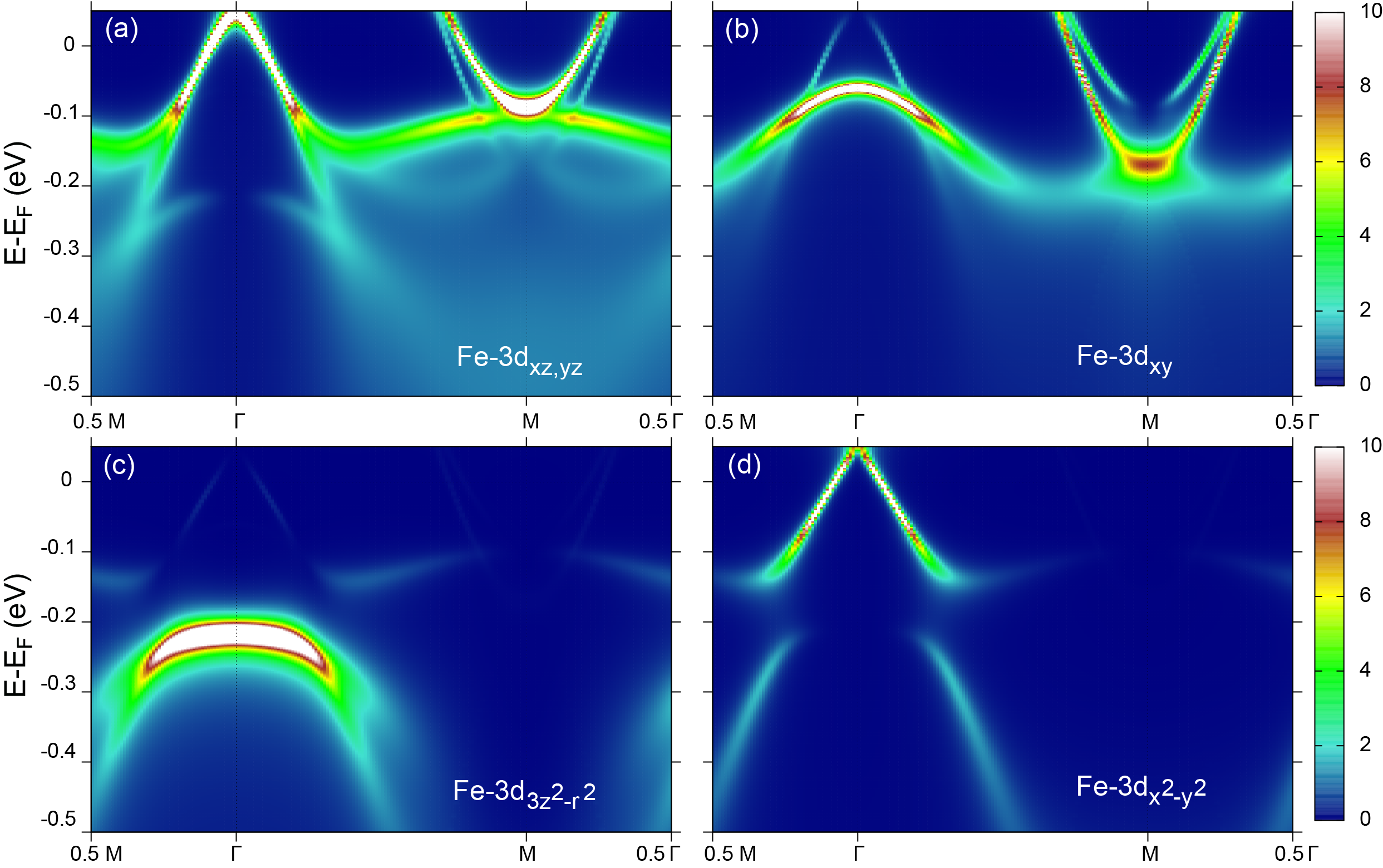}
\includegraphics[width=.58\linewidth]{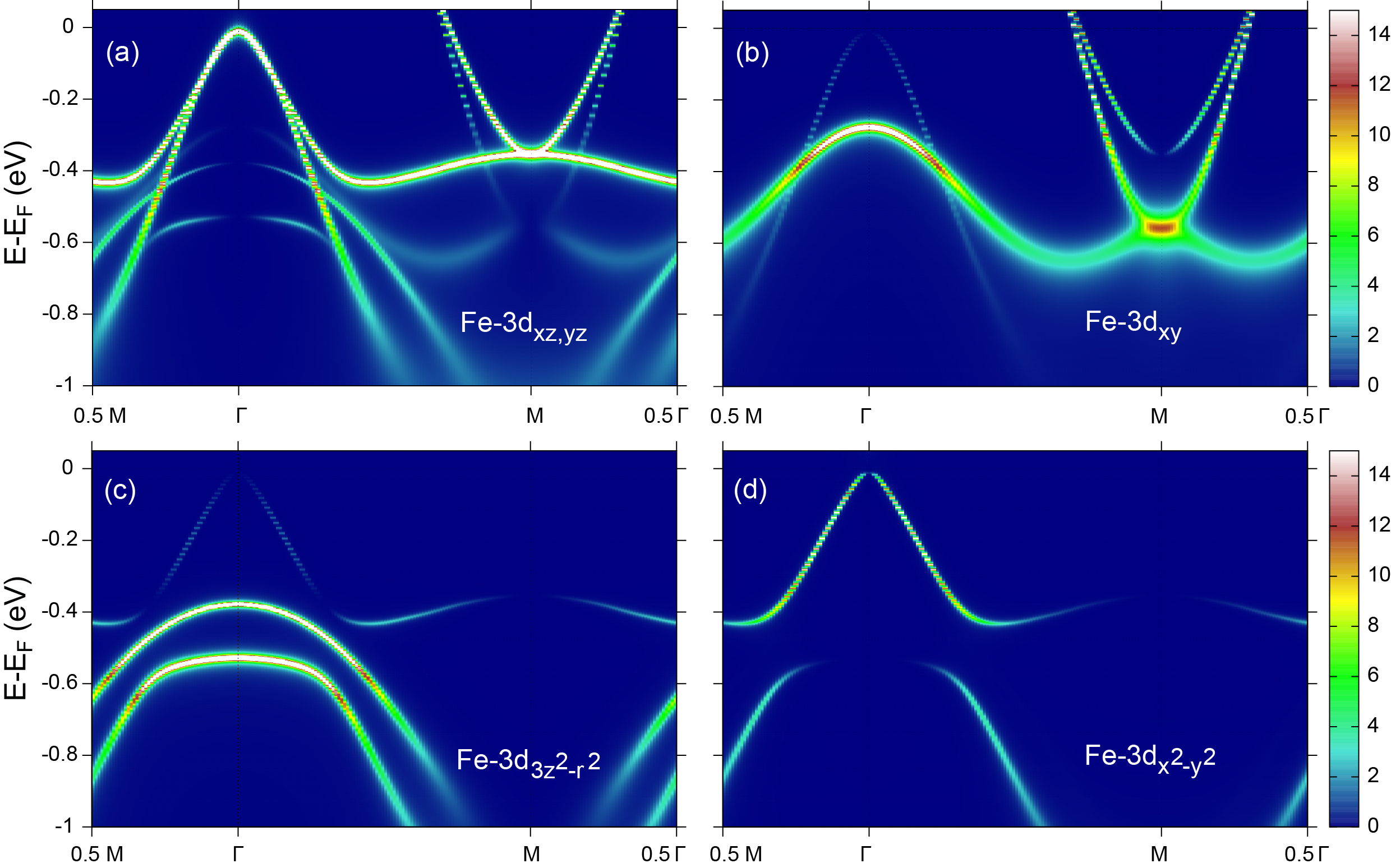}}
\caption{Fig.~S I. LDA+DMFT spectral function map for different Fe-3d orbitals of
FeSe monolayer on SrTiO$_3$ substrate (top) and isolated FeSe monolayer (bottom):
(a) -- Fe-3d$_{xz}$ and Fe-3d$_{yz}$, (b) -- Fe-3d$_{xy}$,
(c) -- Fe-3d$_{3z^2-r^2}$, (d) -- Fe-3d$_{x^2-y^2}$.
Fermi level is at zero energy.}
\end{figure*}
\begin{figure*}[!ht]
\center{\includegraphics[width=.58\linewidth]{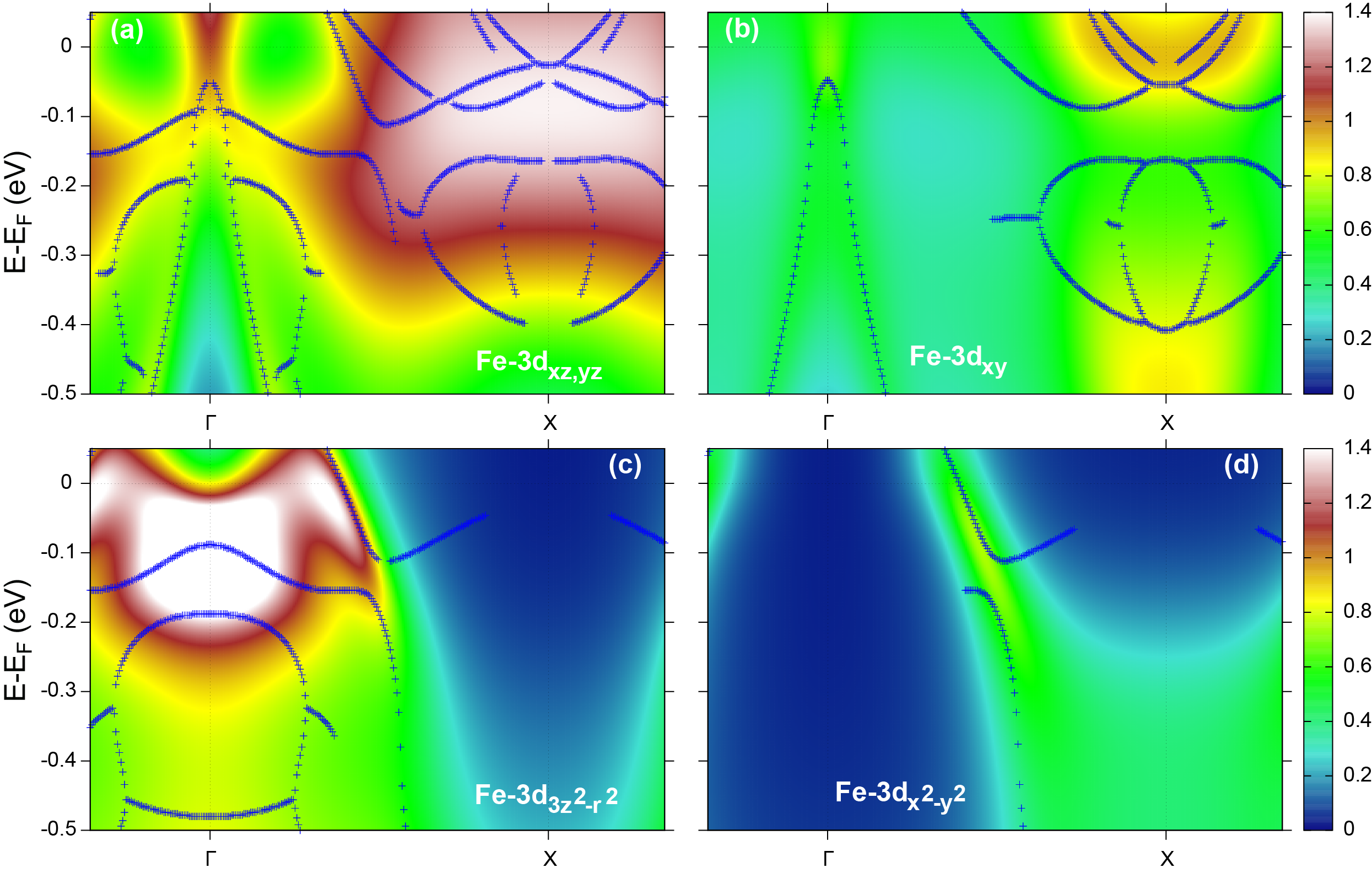}}
\caption{Fig.~S II. LDA$'$+DMFT spectral function map for different Fe-3d orbitals of
K$_{0.76}$Fe$_{1.72}$Se$_2$: (a) -- Fe-3d$_{xz}$ and Fe-3d$_{yz}$,
(b) -- Fe-3d$_{xy}$, (c) -- Fe-3d$_{3z^2-r^2}$, (d) -- Fe-3d$_{x^2-y^2}$.
Maxima of the spectral density are shown with crosses.
Fermi level is zero energy.}
\end{figure*}

\end{document}